\documentclass[prd,reprint,superscriptaddress,nofootinbib,longbibliography,aps]{revtex4-1}
\usepackage[utf8]{inputenc}
\usepackage{graphicx}
\usepackage{longtable}
\usepackage{amsmath}
\usepackage{color}
\usepackage{amssymb}
\usepackage{subfigure}
\usepackage{tabularx}
\usepackage{microtype}
 \usepackage[linktoc=all]{hyperref}
 \usepackage{cleveref}
\usepackage{stackengine}
\usepackage[normalem]{ulem}
\usepackage{slashed}
\usepackage{color}
\usepackage{bm}
\usepackage{braket}
\usepackage{slashed}
\usepackage{comment}
\definecolor{myblue}{rgb}{ 0.188, 0.478,0.858}

\newcommand{\rd}{{\rm d}}

\newcommand{\rms}{\text{rms}}

\newcommand{\vev}[1]{\bigl\langle#1\bigr\rangle}

\begin{document}
\title{Probing Parity Violation in the Stochastic Gravitational Wave Background with Astrometry}

\author{Qiuyue Liang}
\email{qiuyue.liang@ipmu.jp}
\affiliation{Kavli Institute for the Physics and Mathematics of the Universe (WPI), University of Tokyo, Kashiwa 277-8583, Japan}
\affiliation{Center for Particle Cosmology, Department of Physics and Astronomy, University of Pennsylvania, Philadelphia, Pennsylvania 19104, USA}
\author{Meng-Xiang Lin}
\email{mxlin@sas.upenn.edu}
\affiliation{Center for Particle Cosmology, Department of Physics and Astronomy, University of Pennsylvania, Philadelphia, Pennsylvania 19104, USA}
\author{Mark Trodden} 
\email{trodden@upenn.edu}
\affiliation{Center for Particle Cosmology, Department of Physics and Astronomy, University of Pennsylvania, Philadelphia, Pennsylvania 19104, USA}
\author{Sam S. C. Wong}
\email{samwong@cityu.edu.hk}
\affiliation{Center for Particle Cosmology, Department of Physics and Astronomy, University of Pennsylvania, Philadelphia, Pennsylvania 19104, USA}
\affiliation{Department of Physics, City University of Hong Kong, Tat Chee Avenue, Kowloon, Hong Kong SAR, China}

\date{\today}

\begin{abstract}
Astrometry holds the potential for testing fundamental physics through the effects of the Stochastic Gravitational Wave Background (SGWB) in the $\sim 1-100$ nHz frequency band on precision measurements of stellar positions. 
Such measurements are complementary to tests made possible by the detection of the SGWB using Pulsar Timing Arrays. 
Here, the feasibility of using astrometry for the identification of parity-violating signals within the SGWB is investigated. 
This is achieved by defining and quantifying a non-vanishing $EB$ correlation function within astrometric correlation functions, and investigating how one might estimate the detectability of such signals. 
\end{abstract} 

\maketitle

\section{Introduction} 
The totality of gravitational radiation originating from all sources, encompassing both astrophysical \cite{Rajagopal:1994zj,Buonanno:2004tp,Rosado:2012bk,Hils:1990vc,LIGOScientific:2016fpe,Burke-Spolaor:2018bvk} and cosmological phenomena \cite{Hogan:1986qda,Guth:1980zm,Grishchuk:1993te,Kibble:1976sj,Kuroyanagi:2012wm}, 
should give rise to a {\it stochastic gravitational wave background} (SGWB) immersing our galaxy in a sea of gravitational waves (see \cite{Christensen:1992wi} for a review).  
Numerous advanced gravitational wave detectors have the potential to set limits on the amplitude of this background across different frequency ranges~\cite{Lasky:2015lej,Coughlin:2014xua,Pagano:2015hma,Christensen:2018iqi,LIGOScientific:2014sej,Christensen:1992wi,Thrane:2013oya,LIGOScientific:2016fpe,Chaibi:2016dze,Holometer:2016ipr,Jenet:2006sv,Book:2010pf,Fedderke:2021kuy,Wang:2020pmf,Wang:2022sxn,Pardo:2023cag}. Perhaps most exciting is the recently claimed detection of the SGWB in the nanohertz band by pulsar timing array (PTA) collaborations  \cite{Xu:2023wog,NANOGrav:2023gor,EPTA:2023fyk,Reardon:2023gzh}. An important effect of the SGWB is to modulate both the arrival times and directions of photons arriving at Earth from various sources. For PTA measurements, the relevant sources are pulsars and we measure the residuals of the photon arrival times. However, if an SGWB indeed permeates our galaxy, it should also influence astrometry by subtly altering the deflection angles of stars themselves \cite{Book:2010pf,Gwinn:1996gv,Titov:2010zn,Klioner:2017asb,Moore:2017ity,Pardo:2023cag,Wang:2020pmf,Wang:2022sxn,Golat:2022hjf}. 
This is particularly important at the present epoch, when precision measurements of celestial body positions within our galaxy have begun through sensitive astrometry methods. 
Current projects such as the Very Long Baseline Array (VLBA) \cite{Truebenbach:2017nhp,Darling:2018hmc} and Gaia~\cite{Moore:2017ity,Klioner:2017asb}, and future surveys like Theia~\cite{Theia:2017xtk,Malbet:2022lll}, are sensitive to SGWB signals roughly in the $\sim 10^{-9}-10^{-7}\,{\rm Hz}$ band, approximately determined by the inverse of the proposed observing time.
Indeed, the idea that Gaia DR3 data might be used to constrain the amplitude of the SGWB has already been explored \cite{Jaraba:2023djs}.

These novel ways of making gravitational wave measurements raise the possibility of entirely new tests of fundamental physics. Of particular interest in this paper are tests of parity. The prospect of parity-violation in the gravitational sector has long been a topic of theoretical interest. A parity-violating gravitational wave background could arise either from direct violations within the gravitational sector \cite{Lue:1998mq,Jackiw:2003pm,Alexander:2007vt,Satoh:2007gn,Alexander:2009tp,Zhu:2013fja,Zhao:2019xmm,Qiao:2022mln,Jenks:2023pmk} or could be indirectly induced from parity violation in the matter sector \cite{Cook:2011hg,Garcia-Bellido:2016dkw,Obata:2016oym,Obata:2016tmo,Thorne:2017jft,Adshead:2018doq,Adshead:2019igv,Campeti:2020xwn,Unal:2023srk,Niu:2022quw,Niu:2023bsr,Bastero-Gil:2022fme}. 
However, when isotropy is preserved, it turns out that PTA measurements do not respond to the B-mode of the gravitational wave background \cite{Gair:2014rwa,Qin:2020hfy,Liang:2023ary}, 
and hence are not sensitive to polarization information \cite{Kato:2015bye} or to possible parity violations in the SGWB.   
One way forward is to consider relaxing the assumption of isotropy~\cite{Kato:2015bye,Sato-Polito:2021efu,Belgacem:2020nda} although these potential signals can be quite challenging to detect~\cite{Mingarelli:2013dsa,Ali-Haimoud:2020ozu,Ali-Haimoud:2020iyz}. It is therefore worthwhile to consider how signatures of parity violation may manifest themselves in other SGWB surveys. 

In this article we explore the potential of astrometry to identify parity-violating signals in the nanohertz band of the SGWB. By decomposing the deflection vectors of individual stars into vector spherical harmonics, we reveal a non-vanishing $EB$ correlation in the two-point correlation function of stellar position deflections, assuming the presence of an isotropic SGWB with a parity-violation signal. We also carry out initial estimates to assess the detectability of this signal to understand the extent to which it might be used to impose constraints on the size of new parity-violating physics.

The paper is organized as follows. In Sec.~\ref{sec:background}, we briefly review the basics of astrometry measurements of the SGWB. In Sec.~\ref{sec:parity}, we then study the parity violation receiving function and calculate the spectrum relevant for astrometry detection in the spherical harmonic basis. We carry out an estimate of the detectability of such a signal in Sec.~\ref{sec:detectability}, and conclude in Sec.~\ref{sec:conclude}. Throughout the paper we use the metric signature $(-,+,+,+)$. We use $\vec k$ to represent the spatial momentum vector, $k$ stands for the magnitude of the spatial momentum, and $\hat k \equiv \vec k/ k$ represents the unit vector. 

\section{Background}
\label{sec:background} 
The power of astrometry is, of course, the ability to make unprecedentedly precise measurements of stellar positions. We can use these as a tool to measure the SGWB by monitoring the deflection vectors of photons emitted from the stars as gravitational waves distort the spacetime between them and us. For each pair of stars, we can compute the two-point correlation function of their deflection vectors, and from this can extract information about the SGWB. 

The fluctuation of the direction of light from distant stars observed by an inertial observer due to a gravitational plane wave $h_{ij}(t,x)=h_{ij}(\vec{k}) e^{i( \vec{k} \cdot \vec{x}- kt) }$ can be expressed as~\footnote{Throughout this paper we will only focus on the case where the stellar distance is much larger than the wavelength of the gravitational waves. } \cite{Book:2010pf}
\begin{equation}
 \delta n^I (t, \hat n) =e^{I}_{\;\;\mu}  \delta n^\mu (t, \hat n) = {\cal R}^{IJK}(\hat{n}, \hat{k})  h_{JK} (\vec{k} ) e^{-i k t}\ , 
\end{equation}
with
\begin{equation}
 {\cal R}^{IJK}(\hat{n}, \hat{k})  =  \frac{\hat n^I + \hat{k}^I}{2 (1 + \hat{k}\cdot\hat{n} )}\hat n^J \hat n^K  - \frac{1}{2} \delta^{IJ}\hat n^K  \ ,
\end{equation}
where $\hat{n}$ and $\hat{k}$ are the line of sight unit vector from the observer to the star, and the wave vector respectively. Note that we have expressed the deflection vector $\delta n^\mu$ of the photon in the non-spinning inertial frame of a local observer via the basis $e^I_{\;\; \mu}$. 
One can view the 3-index quantity $R^{IJK}$ as an analogy of the receiving function in the gravitational wave literature\cite{Thrane:2013oya,Isi:2018miq}, however with an extra index representing the vector nature of the observable. The next step is to study the correlated signatures of the deflection vectors among the sources. 

\section{Parity violation}
\label{sec:parity}
Under the standard mode expansion for plane gravitational waves
\begin{align}
    h_{ij}(t, \vec{x}) = \sum_{S=+,\times} \int \frac{\rd^3 {\vec k}}{(2\pi)^3} \; h_S(t, \vec{k})e^S_{ij}(\hat k) e^{  i  \vec{k}\cdot \vec x }  \,,
\end{align}
a stochastic gravitational wave background is characterized by the two-point functions~\footnote{Note that we have used a convention that agrees with that used for the in-in correlators used in the computation of inflationary correlation functions.}
\begin{align}
   &\vev{h_S(t, \vec{k}) h_{S'}(t, \vec{k}')} = (2\pi)^3 \delta(\vec{k}+ \vec{k}')P_{S S'}(k) ,\nonumber  \\
  &[P_{S S'}(k)] = \begin{bmatrix}  P_{++}(k) & P_{+\times}(k) \\
     P_{\times +}(k)  & P_{\times \times}(k) 
   \end{bmatrix} ,
\end{align}
where we have assumed translational and rotational invariance so that $P_{++}=P_{\times\times}$. Within the galaxy, flat spacetime is a good approximation, and therefore equal time correlators are time-independent.  The two-point correlation function for $\delta n^I$ is then given by 

\begin{eqnarray}
\begin{aligned}
     \vev{\delta n^I (t, \hat n) & \delta n^J (t, \hat n')} \nonumber\\
    =&\sum_{S, S'} \int \frac{\rd^3 {\vec k}}{(2\pi)^3} {\cal R}^{IKL}(\hat{n}, \hat{k})  {\cal R}^{JMN}(\hat{n}',   \hat{k}) \nonumber\\
    \times & e^S_{KL}(\hat k) e^{S'}_{MN}(-\hat k) P_{SS'}(k) \nonumber \\
    = & \sum_{S, S'} \int_0^{\infty} \rd k  \frac{k^2}{(2\pi)^3} P_{SS'}(k) H^{IJ}_{SS'}(\hat{n},\hat{n}') \ ,
\end{aligned}       
\end{eqnarray}
where 
\begin{align} \label{eqn:Hoverlred}
    H^{IJ}_{SS'}(\hat{n},\hat{n}') &= \int \rd^2 \Omega_{\hat{k}} {\cal R}^{IKL}(\hat{n}, \hat{k})  {\cal R}^{JMN}(\hat{n}', -\hat{k}) \nonumber\\
    & \times e^S_{KL}(\hat k) e^{S'}_{MN}(-\hat k) \ .
\end{align}  
 
It is simple to show that the definition of $\delta n^I$ implies that $H_{SS'}^{IJ}$ should satisfy 
\begin{align}
    \hat n_{I}H_{SS'}^{IJ}(\hat{n},\hat{n}') =  \hat n'_{J}H_{SS'}^{IJ}(\hat{n},\hat{n}') =0 \ .
\end{align}
Writing $\hat{k} = (\sin\theta \cos\phi, \, \sin\theta \sin\phi,\, \cos\theta)$ in Cartesian components, we employ the convention
\begin{align}
    e^+_{IJ} = \hat{\phi}_I \hat{\phi}_J - \hat{\theta}_I\hat{\theta}_J,\quad    e^{\times}_{IJ} = \hat{\phi}_I \hat{\theta}_J + \hat{\theta}_I \hat{\phi}_J \ ,
\end{align}
where 
\begin{align}
    &\hat{\phi} = ( -\sin\phi ,\, \cos \phi, \,0)\ ,  \nonumber\\
    &\hat{\theta}= (\cos \theta \cos\phi, \,\cos\theta \sin\phi,\, -\sin\theta)\ .
\end{align}
Under a parity transformation $\theta \to \pi - \theta $, $\phi \to \phi+\pi$, and $\hat{k} \to -\hat{k} $, one finds 
\begin{equation*}
    e^+_{IJ} \to e^+_{IJ}\ , \quad e^{\times}_{IJ} \to - e^{\times}_{IJ} \ .
\end{equation*}
Therefore from the definition \eqref{eqn:Hoverlred}, $H^{IJ}_{++}(\hat{n}, \hat{n}')$ and $H^{IJ}_{\times\times}(\hat{n}, \hat{n}')$ are parity even, while  $H^{IJ}_{+\times}(\hat{n},\hat{n}')$ and $H^{IJ}_{\times+}(\hat{n}, \hat{n}')$ are parity odd under $\hat{n} \to -\hat{n}$,  $\hat{n}' \to -\hat{n}'$.

\subsection{The overlap reduction function $H_{+\times}^{IJ}(\hat{n}, \hat{n}')$}

\begin{figure}
\centering
\includegraphics[scale=0.6]{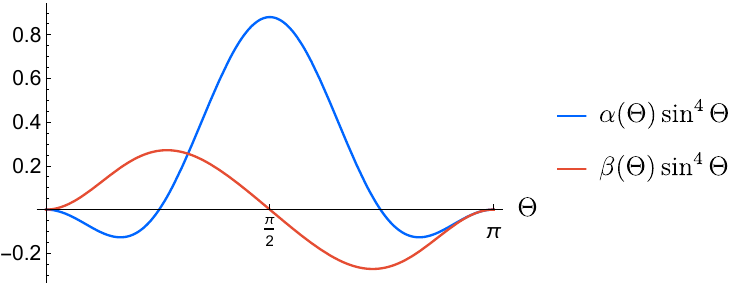} \includegraphics[scale=0.6]{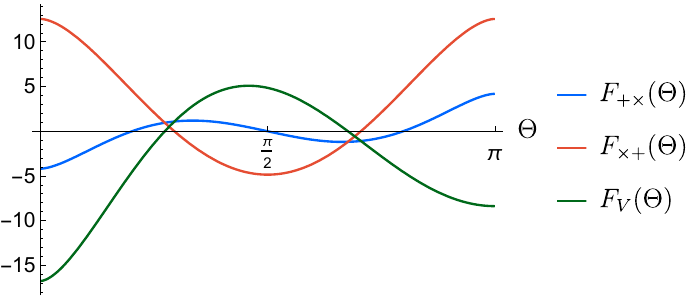} 
\caption{Top: the functions $ \alpha( \Theta)\sin^4\Theta $ and $ \beta( \Theta)\sin^4\Theta $ defined in Eq.\eqref{eq,alphabeta}; bottom:  $F_{+\times(\times+)}(\Theta)$ and $F^V(\Theta)$ defined in Eq.\eqref{eq,FEB}.}
\label{fig:alphabeta}
\end{figure}

The symmetry properties of $H_{+ \times (\times +)}^{IJ}$ discussed above imply that, given any orientation of $\hat{n}$ and $\hat{n}'$, they can be written as 
\begin{align}
    H_{+ \times}^{IJ}(\hat{n}, \hat{n}') =  \alpha(\Theta) A^I B_2^J +  \beta(\Theta) B_1^I A^J \ , \\
    H_{ \times +}^{IJ}(\hat{n}, \hat{n}') =  \alpha_2(\Theta) A^I B_2^J +  \beta_2(\Theta) B_1^I A^J \, ,
\end{align}
where 
\begin{align}
    \vec{A} = \hat{n} \times \hat{n}', \quad \vec{B}_1 = \hat{n} \times \vec{A}, \quad \vec{B}_2 = \hat{n}' \times \vec{A} .
\end{align}
We can then extract the functions $\alpha(\Theta)$ and $\beta(\Theta)$ using
\begin{align}
     A_I B_{2J}H_{+ \times}^{IJ}(\hat{n}, \hat{n}') = \alpha(\Theta) (w^2-1)^2 \, \nonumber
\end{align}
\begin{align}
     B_{1I} A_{J}H_{+ \times}^{IJ}(\hat{n}, \hat{n}') = \beta(\Theta) (w^2-1)^2 \ ,
\end{align}
where $w=\hat{n}\cdot\hat{n}'=\cos\Theta$. A direct evaluation of the integral \eqref{eqn:Hoverlred} gives
\begin{widetext}
\begin{align}\label{eq,alphabeta}
 \alpha_2(\Theta)  &= \beta(\Theta), \quad  \beta_2(\Theta) = \alpha(\Theta), \nonumber \\
    \alpha(\Theta) &= -\frac{\pi \left[(5+18 \log 2) w^2 -5+\log 64+3 (w-1)^3 \log (1-w )-3 (w +1)^3 \log (1+w)\right] }{3 (w^2-1)^2}  \ ,\nonumber \\
    \beta(\Theta) &=  -\frac{\pi \left[ (7+\log 64)w^3-\left(7-18 \log 2\right)w-3 (w-1)^3 \log (1-w)-3 (w+1)^3 \log (1+w)\right]}{3 (w^2-1)^2} \ .
\end{align}
\end{widetext}
We plot these in the top panel of Fig.~\ref{fig:alphabeta}.

With these tools in hand, we can use $\vev{\delta n^I (t, \hat n) \delta n^J (t, \hat n')} $ to construct scalar quantities that contain information about parity violation. There are in principle infinitely many such quantities, constructed by applying ever more powers of derivatives. For example, at the lowest order in derivatives,
\begin{align}
  \epsilon_{IJK}\hat n^I  \vev{\delta n^J (t, \hat n) \delta n^K (t, \hat n')}
\end{align}
contains contributions from the parity violating components $H_{+\times}$ and $H_{\times +}$. However, the parity conserving components, $H_{++}$ and $H_{\times \times}$, also enter this expression. At the next order in the number of derivatives, we have
\begin{align}
\label{eq,scalarquantity}
    &\quad \epsilon_{JKL}\nabla_I\nabla'{}^{L} \left(\hat n'{}^K\vev{\delta n^I (t, \hat n) \delta n^J (t, \hat n')} \right)  \nonumber \\
   & = {\cal A}_{+\times}  F_{+\times}(\Theta)+{\cal A}_{\times +}F_{\times +} (\Theta) \ ,
\end{align}
where $\nabla'$ is the covariant derivative acting on $\vec x'$, 
\begin{align} \label{eqn:Fdef}
    F_{SS'} = \epsilon_{JKL}\nabla_I\nabla'{}^{L} \left(\hat n'{}^K H^{IJ}_{SS'} \right)  ,
\end{align}
and
\begin{align} \label{eq,amplitudess}
     {\cal A_{SS'}}  = \int_{k_{\rm min}}^{k_{\rm max}} \rd k  \frac{k^2}{(2\pi)^3} P_{SS'}(k) \ .
\end{align}
This quantity is non-zero {\it only if} the theory is parity violating.
When focusing on an isotropic background \cite{Kato:2015bye}, we always have $P_{+\times} = -P_{\times+}$. It is therefore valid to take ${\cal A}_{+\times}=-{\cal A}_{\times+}$, in which case we may express Eq.\eqref{eq,scalarquantity} as
\begin{align} \label{eq,FEB}
    \epsilon_{JKL}\nabla_I\nabla'{}^{L} \left(n'{}^K\vev{\delta n^I (t, \hat n) \delta n^J (t, \hat n')} \right)
    \equiv{\cal A}_{+\times} F^V \ ,
\end{align}  
where $F^V = F_{+\times} -F_{\times +} $.

We will see in the following section that this quantity is related to $EB$ correlations when $\delta n^I$ is decomposed into vector spherical harmonics.

\subsection{Spectrum in spherical harmonic basis}
Since $  \vev{\delta n^I (t,\hat n) \delta n^J (t, \hat n')}$ is a correlation function on the two sphere and  $\delta  n^I$ is perpendicular to $\hat{n}$, it is natural and useful to express $\delta n(t, \hat{n})$ in a basis of vector spherical harmonics 
\begin{align}
    \delta n (t, \hat{n}) = \sum_{\ell m} \left[ \delta n_{E\ell m}(t) \vec{Y}^E_{\ell m}(\hat{n}) +  \delta n_{B\ell m}(t) \vec{Y}^B_{\ell m}(\hat{n})   \right] \ ,\nonumber
\end{align}
where
\begin{align}
    &\vec{Y}^E_{\ell m}(\hat{n}) = \frac{1}{\sqrt{\ell(\ell+1)}} \nabla Y_{\ell m}(\hat{n})\ , \nonumber \\
   & \vec{Y}^B_{\ell m}(\hat{n}) = \frac{1}{\sqrt{\ell(\ell+1)}} \hat{n}\times \nabla Y_{\ell m}(\hat{n}) \ , \nonumber \\
   &  \int \rd^2 \Omega_{\hat{n}} \delta^{IJ} Y^Q_{\ell m I} (\hat{n}) Y^{Q' *}_{\ell' m' J} (\hat{n}) = \delta_{Q Q'}\delta_{\ell \ell'} \delta_{mm'} \ . 
\end{align}
The $EB$ correlation function then takes the form
\begin{align}
    &\quad  \vev{\delta n_{E \ell m}(t)\delta n_{B \ell' m'}(t)^*  } \nonumber\\
   & = \int \rd^2 \Omega_{\hat{n}}  \rd^2 \Omega_{\hat{n}'} Y^{E*}_{\ell m I}(\hat{n})  Y^{B}_{\ell' m' J}(\hat{n}')   \vev{\delta n^I (t, \hat n) \delta n^J (t, \hat n')} \ .
\end{align}
Since we have $\vec{Y}^E_{\ell m}(-{\hat n}) = (-1)^{\ell +1} \vec{Y}^E_{\ell m}({\hat n}) $ and $\vec{Y}^B_{\ell m} = (-1)^{\ell } \vec{Y}^B_{\ell m}({\hat n})$, only $H^{IJ}_{+\times}(\hat{n}, \hat{n}')$ contributes when $\ell$ is odd and only $H^{IJ}_{\times +}(\hat{n}, \hat{n}')$ contributes when $\ell$ is even. Also, because $\delta n^I$ has no radial component, we may integrate by parts to obtain
\begin{align} \label{eqn:EBcorr}
    \quad  \vev{\delta n_{E \ell m}(t) & \delta n_{B \ell' m'}(t)^*  } \nonumber\\
   & = \frac{1 }{\ell(\ell+1)} \int \rd^2 \Omega_{\hat{n}}  \rd^2 \Omega_{\hat{n}'} Y^{*}_{\ell m }(\hat{n})  Y^{}_{\ell' m' }(\hat{n}') \nonumber\\
    &\qquad  \epsilon_{JKL}\nabla_I\nabla'{}^{L} \left(\hat n'{}^K\vev{\delta n^I (t, \hat n) \delta n^J (t, \hat n')} \right)  \nonumber \\
    &= \frac{1 }{\ell(\ell+1)} \int \rd^2 \Omega_{\hat{n}}  \rd^2 \Omega_{\hat{n}'} Y^{*}_{\ell m }(\hat{n})  Y^{}_{\ell' m' }(\hat{n}') \nonumber \\
    & \quad ({\cal A}_{+\times} F_{+\times }(\Theta)+{\cal A}_{\times+} F_{\times+}(\Theta))\ ,
\end{align}
where the amplitude is defined through Eq.\eqref{eq,amplitudess}.  
An explicit computation of \eqref{eqn:Fdef} gives
\begin{align}
 F_{+\times} (\Theta)& =\frac{4}{3} \pi   \Big(-w (1+\log 64 )+3 (w-1) \log (1-w) \nonumber \\
&\qquad \qquad +3 (w+1) \log (w+1) \Big) \ , \nonumber\\
F_{\times + } (\Theta)& =4 \pi   \Big( 1- \log 4 + (1-w) \log (1-w)\nonumber \\
& \qquad \qquad  +(1+w) \log(1+w) \Big)  \ , 
\end{align}
where, again, $w = \cos(\Theta)$.  
We plot these in the bottom panel of Fig.~\ref{fig:alphabeta}.

We can further expand the function $F_{SS'}(\Theta) $ in terms of Legendre polynomials, 
\begin{eqnarray}
    F_{SS'}(\Theta) &=& \sum_\ell  F_{SS'}^\ell  P_\ell (\cos\Theta) \nonumber\\
    &=& \sum_{\ell  m} \frac{4\pi }{2\ell +1} F_{SS'}^\ell Y_{\ell m}(\hat n)Y^*_{\ell m}(\hat n')\ ,
\end{eqnarray}
so that equation \eqref{eqn:EBcorr} simplifies to
\begin{align}  \label{eq,ClEB}
  &\quad  \vev{\delta n_{E \ell m}(t)\delta n_{B \ell' m'}(t)^*  } \nonumber \\
   &= \frac{\delta_{\ell \ell'} \delta_{m m'}}{\ell(\ell+1)} \frac{4\pi  }{2\ell +1}  \left( {\cal A}_{+\times}F_{+\times}^\ell  + {\cal A}_{\times+}F_{\times + }^\ell \right) \nonumber\\
   & =   \frac{\delta_{\ell \ell'} \delta_{m m'}}{\ell(\ell+1)} \frac{4\pi {\cal A}_{+\times} }{2\ell +1} F_V^\ell \ .
\end{align}
Here we have assumed ${\cal A}_{+\times} = -{\cal A}_{\times+}$ in the second equality. Again, $F_{+\times}$ ($F_{\times+}$) contributes only when $\ell$ is odd (even). This formalism has previously been used~\cite{Kato:2015bye,OBeirne:2018slh} to explore polarization information. This approach can be generalized to an anisotropic SGWB by allowing for the possibility that ${\cal A}_{+\times} \neq -{\cal A}_{\times+}$.

\begin{figure}[h!]
\centering
\includegraphics[scale=0.6]{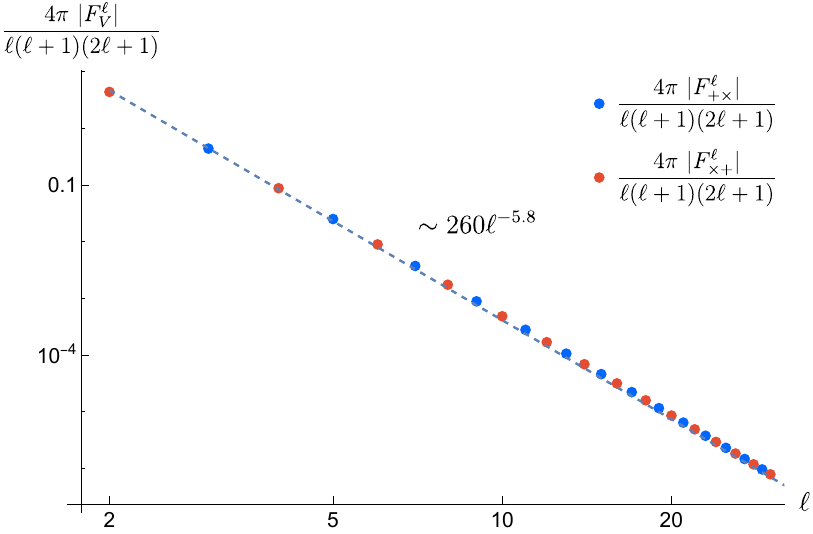} 
 \caption{ The log-log plot of the angular power spectrum Eq.\eqref{eq,ClEB} for the first thirty multipoles. The red dots are the coefficients obtained from $F_{\times+}$, and only contribute to the even modes. The blue dots are those from $F_{+\times}$ and only contribute to the odd modes. One can see that the angular power spectrum falls roughly as $l^{-5.8}$. We provide the exact numbers in Table \eqref{tab:Fell} below.}
\label{fig:background}
\end{figure}

In Fig~\ref{fig:background} we provide numerical results for the angular power spectrum Eq.\eqref{eq,ClEB} for the first 30 multipole modes. One can see that $F_{+\times}^\ell$ ($F_{\times+}^\ell$) vanishes for $\ell$ even (odd), as expected.  It is worth noticing that the coefficients of both $F_{+\times}^\ell$ and $F_{\times+}^\ell$ fall off approximately as $\ell^{-6}$, which is also the same decaying behavior as $(\nabla_I \nabla_J H^{IJ}_{++(\times\times)})_\ell$ \cite{Book:2010pf}. We provide the exact values of these quantities in Table \ref{tab:Fell} below.

\begin{table*}[!] 
\centering
 \begin{tabular}{|c|c c c c c c c c c c|} 
 \hline 
 $\ell$ & 1 & 3 & 5 & 7 & 9 & 11 & 13 & 15 & 17 & 19 \\
 \hline 
 $\frac{4\pi}{\ell(\ell+1)(2\ell+1)}F_{+\times}^\ell$ & 0  & $-\frac{2 \pi ^2}{45}$  & $-\frac{4 \pi ^2}{1575}$ & $-\frac{\pi ^2}{2646}$ & $-\frac{2 \pi ^2}{22275}$ & $-\frac{2 \pi ^2}{70785}$ &$ -\frac{4 \pi ^2}{372645}$ & $-\frac{\pi ^2}{214200}$ & $-\frac{\pi ^2}{444771}$ & $-\frac{2 \pi ^2}{1705725}$ \\
 \hline
 \end{tabular}
 \begin{tabular}{|c|c c c c c c c c c c|} 
 \hline 
 $\ell$ & 2 & 4 & 6 & 8 & 10 & 12 & 14 & 16 & 18 & 20 \\
 \hline 
 $\frac{4\pi}{\ell(\ell+1)(2\ell+1)}F_{\times+}^\ell$ & $ \frac{4 \pi ^2}{9}$  & $\frac{2 \pi ^2}{225}$ &  $\frac{2 \pi ^2}{2205}$ & $\frac{\pi ^2}{5670}$ & $\frac{4 \pi ^2}{81675}$ & $\frac{2 \pi ^2}{117117}$ & $\frac{\pi ^2}{143325}$ & $\frac{\pi ^2}{312120}$ & $\frac{4 \pi ^2}{2485485}$ & $\frac{2 \pi ^2}{2304225}$ \\
 \hline
 \end{tabular}
\caption{The table of coefficients of $F_{\times+}^\ell$ and $F_{+\times}^\ell$ defined in \eqref{eq,ClEB}.
} \label{tab:Fell}
\end{table*}

\section{Detectability}
\label{sec:detectability}
We now turn briefly to the possibility of detecting the parity-violating signatures in gravitational wave signals using astrometry with current and future telescopes.

When a light ray passes through a SGWB, the root-mean-square (rms) deflection angle caused by the underlying metric perturbation is (\cite{Book:2010pf,Braginsky:1989pv,Kaiser:1996wk}) proportional to the gravitational wave strain amplitude, $\delta_\rms \sim h_\rms $. This deflection can be interpreted as a measure of the angular velocity (or proper motion) of the light source via $\omega_\rms \sim f h_\rms$, which, for $N$ sources, has a correlated signal of order 
\begin{equation}
    f h_\rms \sim \Delta \theta /\left(T \sqrt{N}\right) \ , 
\end{equation}where $\Delta\theta$ is the angular resolution, and $T$ is the observation time. 
Considering the current Gaia catalog, for an optimistic estimate we take the resolution to be the ideal case $\Delta\theta\sim 10\mu{\rm as}$ and the number of sources to be $N\sim 10^8$ for stars and $N\sim 10^6$ for quasars~\cite{vallenari2023gaia}. The observation time is $T\sim 3{\rm yr}$ and the most sensitive frequency is $f\sim1/T$. Therefore, we obtain the optimistic estimate $h_\rms \sim 5\times10^{-15}$ for stars and $h_\rms \sim 5\times10^{-14}$ for quasars. However, not all sources will achieve the ideal resolution and proper motions will further contaminate the data \cite{Jaraba:2023djs}. As a result, the realistic Gaia constraining power could be one or two orders of magnitude below the signal extrapolated from the current NANOGrav 15 yr results~\cite{NANOGrav:2023gor}, $ h_{\rm signal}(f=1/3\,{\rm yr}^{-1}) \sim  10^{-15}$. Future experiments, such as Theia~\cite{Jaraba:2023djs,Garcia-Bellido:2021zgu}, may provide constraining power closer to the optimistic estimate.

These estimates also apply to parity violating correlations. 
It is therefore worth considering how we might specifically minimize the noise in measurements of the $EB$ correlation signal on which we have focused in this paper. One way to do this might be by using quasars rather than stars as the observational sources. The global motions of stars can induce large-scale correlations and can generate both E-mode and B-mode signals. Since all of the observed stars are in one galaxy (our Milky Way), we cannot expect their $EB$ correlations to be suppressed just because the underlying theory is parity conserving. However, for quasars, the primary source of noise in large-scale correlations of changes in their positions is that induced by cosmological large-scale structure, which does not produce an $EB$ correlation due to its inherent parity-conservation.\footnote{ On cosmological scales, the power spectrum might take a different form compared to that on galactic scales, and the quasar term might not be negligible. This possibility merits further study.}  Hence, as we look to the future, precision measurements of quasars might be an even more interesting target in the quest to find $EB$ correlations. The analysis in this paper is easily adapted to that case.

\section{Conclusions and Discussions}\label{sec:conclude}
In this paper we have discussed the possibility of using astrometry to detect parity-violating signals in the SGWB. Specifically, we have calculated the $EB$ correlation function of gravitational-wave-induced stellar position changes, which is non-zero if and only if the underlying theory is not parity conserving. Parity violation can occur in both the matter sector and the gravity sector but is not sensitive to the source of the SGWB.
We have further estimated the possibility of detecting this signal in current and future astrometric surveys, and have argued that, as we look to the future, quasars may ultimately be the best targets for searching the parity-violating signals.

\acknowledgments
We thank Gary Bernstein, Eanna Flanagan, Wayne Hu, and Kris Pardo for useful discussions. The work of QL, MT and SW is supported in part by US Department of Energy (HEP) Award DE-SC0013528. QL is partly supported by World Premier International Research Center Initiative (WPI), MEXT, Japan. M-X. L. is supported by funds provided by the Center for Particle Cosmology.

\bibliographystyle{utphys}
\bibliography{GW.bib} 

\end{document}